\documentclass[11 pt]{article}
\usepackage{graphicx, amssymb, amsmath, epsfig, url, setspace, subfigure, color}
\usepackage{natbib}
\newtheorem{definition}{Definition}

\usepackage{algorithm}
\usepackage{algorithmicx}
\usepackage{algpseudocode}

\newcommand{\bD}{\mathbf{D}}

\begin{document}
\title {Providing Access to Confidential Research Data Through Synthesis and Verification: An Application to Data on Employees of the U.S.\  Federal Government}
\author{Andr\'es F. Barrientos, Alexander Bolton, Tom Balmat, Jerome P. Reiter, John M. de Figueiredo\\Ashwin Machanavajjhala, Yan Chen, Charley Kneifel, Mark DeLong\footnote{Andr\'es F. Barrientos is Postdoctoral Associate, Department of Statistical Science, Duke University, Durham, NC 27708 (email: afb26@stat.duke.edu); Alexander Bolton is Assistant Professor, Department of Political Science, Emory University, Atlanta, GA 30322 (abolton@emory.edu); Tom Balmat is Statistician, Social Science Research Institute, Duke University, Durham, NC 27708 (thomas.balmat@duke.edu); Jerome Reiter is Professor, Department of Statistical Science, Duke University, Durham, NC 27708 (jreiter@duke.edu); John de Figueiredo is Russell M.\ Robinson II Professor, Law School and Fuqua School of Business, Duke University, Durham NC 27708 (jdefig@duke.edu); Ashwin Machanavajjhala is Assistant Professor, Department of Computer Science, Duke University, Durham, NC 27708 (ashwin@cs.duke.edu); Yan Chen is Graduate Student, Department of Computer Science, Duke University, Durham, NC 27708 (yanchen@cs.duke.edu); Charley Kneifel is Senior Technical Director, Office of Information Technology, Duke University, Durham, NC 27701 (charley.kneifel@duke.edu); and, Mark DeLong is Director of Research Computing, Office of Information Technology, Duke University, Durham, NC 27701 (mark.delong@duke.edu).  This research was supported by NSF grants ACI-14-43014 and SES-11-31897, as well as the Alfred P. Sloan Foundation grant G-2-15-20166003.}}


\date{}


\maketitle





\begin{abstract}


Data stewards seeking to provide access to large-scale social science data 
face a difficult challenge. They have to share data in ways that  protect privacy and confidentiality, are informative for many analyses and purposes, and are relatively straightforward to use by data analysts.   One approach suggested in the literature is that data stewards  generate and release  
synthetic data, i.e., data simulated from statistical models, while also providing users access to a verification server that allows them to assess the quality of inferences from the synthetic data.  We present an application of the synthetic data plus verification server approach to longitudinal 
data on employees of the U.\ S.\ federal government.  As part of the application, we present a novel model for generating synthetic career trajectories, as well as strategies for generating high dimensional, longitudinal synthetic datasets.  We also present novel verification algorithms for regression coefficients that satisfy differential privacy.  
We illustrate the integrated use of synthetic data plus verification via analysis of differentials in pay by race. The integrated system performs as intended, allowing users to explore the synthetic data for potential pay differentials and learn through verifications which findings in the synthetic data hold up and which do not.  The analysis on the confidential data reveals pay differentials across races not documented in published studies.  


\end{abstract}


\section{Introduction}\label{intro}

Widespread access to  large-scale social science datasets
greatly enhances 
the work of evidence-based policy makers, social scientists, and statisticians.  
Yet, widespread dissemination of large scale social science data also 
carries a significant social cost: it  puts data subjects' privacy and 
confidentiality at  risk.  Simply stripping unique identifiers
like names and exact addresses, while necessary, generally does not suffice 
to protect confidentiality.  As is well documented
\citep[e.g.,][]{sweeney97, sweeneyhealth, narayanan08:robust, facebook}, ill-intentioned users
may be able to link records in the released data to records
in external files by  matching on variables common to both sources. 
These threats are particularly serious for large-scale social science
data.  Such data often come from administrative or privately collected
sources so that, by definition, someone other than the organization
charged with sharing the data knows the identities and (a large number
of) attributes of data subjects.   
Large-scale social science data also typically include many variables that,
since the data arguably are known by others, could serve as matching
variables.  

As the size, richness, and quality of social science data have increased, so too have the 
threats to confidentiality. Confronted with these risks, responsible data stewards face a
difficult dilemma: how can they provide access to confidential social
science data while protecting confidentiality of data subjects'
identities and sensitive attributes?  Often data stewards---whether
in academia, government, or industry---default 
to restricting access to carefully vetted and approved researchers via licensing arrangements or physical
data enclaves. This is only a partial solution. It
denies the benefits of data access to broad subsets of society including, for example, students who need data for learning the skills 
of data analysis and citizen scientists seeking to understand their society.



One approach  that has been suggested in the literature \citep[e.g., ][]{karr:reiter:14} to deal with this dilemma is for data stewards to use an integrated system  comprising three components, namely (i) a fully synthetic dataset \citep{rubin:1993} intended for wide access, (ii) a verification server \citep{vs08} that allows users to assess the quality of  inferences from the synthetic data, and (iii) means for approved users to access the confidential data, such as by secure remote access or a physical enclave.  We review the rationale for this approach in Section \ref{SynthVerif}.  As far as we are aware, however, no one has implemented or illustrated this type of integrated system  for  complex data typically used in social science research.  

In this article, we present an application of synthesis with verification on longitudinal data comprising the workforce of the United States federal government from 1988 to 2011.  Specifically, we generate an entirely synthetic federal workforce using administrative data from the Office of Personnel Management (OPM).  
The dimensionality and longitudinal structure present many complications for synthetic data generation; indeed, part of the contribution of this article is to describe strategies for generating synthetic datasets with such complexity. Among these strategies is a new Bayesian model for generating career trajectories. We also include assessments of  the analytic validity of and potential disclosure risks in the synthetic data.  To illustrate the benefits and usage of verification servers, we estimate  regression models with the synthetic data  that assess systematic differences in employee salaries by race and gender, and we investigate how such differences change over time.  We verify the results using novel verification measures that we design to satisfy differential privacy \citep{dwork:06}.  
We empirically evaluate the performance of the new differentially private measures, illustrating when to expect them to yield analytically useful results and when not to do so.   
Finally, we validate the regression results using the confidential data.


The findings are remarkable for both methodological and substantive reasons.  For the former, the integrated system performs as advertised, allowing us to see the validity, and shortcomings, of the synthetic data results.  For the latter, the confidential data suggest that, given our model specification, (i) the differential in pay for white and black female employees has been increasing over time, and that (ii) Asian male employees make substantially less on average than white male employees over much of the time frame we analyzed.  As far as we know, neither of these findings have been previously documented in the public sector at this magnitude or detail. 

 
The remainder of this article is organized as follows. In Section \ref{SynthVerif}, we review the general framework of providing synthetic data with verification via an integrated system.   In Section \ref{background}, we describe the OPM data in more detail and outline the procedures used to generate the synthetic data, including the new Bayesian model for generating synthetic careers.  In Section \ref{verification}, we describe the novel differentially private verification measures.   In Section \ref{Analysis}, we mimic a usage of the integrated system to analyze the differential pay gap: we start with the synthetic data, verify findings using the differentially private measures, and repeat the analysis on the confidential data.
Although the big picture concepts underlying this integrated system have been highlighted previously \citep[][p. 57]{callier, ryanmurray}, this analysis represents the first illustration of the full framework on genuine data. The substantive analysis itself is notable, as we are unaware of any recent, public studies by the federal government on racial disparities in pay.  The relative decrease in gains by black women and the relatively stagnant differential between white and black men suggest limited recent progress toward pay equality for these groups.
Finally, in Section \ref{Conclusion}, we discuss implementation issues for these kinds of integrated systems and suggest topics for future research. 

\section{The Framework: Synthesis with Verification}\label{SynthVerif}




We use a framework that integrates three key ideas
from the literature on confidentiality protection and data access. 
The first idea is to provide  synthetic public use files, as proposed by \citet{rubin:1993} and others 
\citep[e.g., ][]{little:1993,fien:94,raghu:rubin:2001, reiter:raghu:07, drechslerbook}. Such files 
comprise individual records with every value replaced
with simulated draws from an estimate of the multivariate distribution of the
confidential data. When generated appropriately, synthetic data can preserve many, but
certainly not all, important  associations in the confidential data. They also should carry low risks of re-identification disclosures, since the released data do not correspond to actual
records. This largely eliminates the kinds of record linkage attacks that have broken typical  disclosure control methods, as it is nonsensical 
for ill-intentioned users 
to match synthetic records to external files.  

While synthetic data have been used to release public use
versions of several high profile social science datasets \citep{abowd06, hawalaacs, onthemap, IABsynthesis, lbdisr}, at present they have a critical
weakness. Users of synthetic data cannot determine how much their 
analysis results have been impacted by the synthesis process if all they have are the synthetic data. 
This limitation leads to the second idea that is integrated in the
framework: provide users access to verification servers \citep{vs08}. 
A verification server is a query-based system that (i)
receives from the user a statistical query that enables comparison of results from the synthetic and 
confidential data, and (ii)  returns an answer to the query without allowing the user to view the confidential data directly \citep{karr:reiter:14}.  With the output from a verification server, users can
decide whether or not results based on the synthetic data are of
satisfactory quality for their particular purposes. Crucially, however, verifications also leak information about the confidential data. 
Thus, we should apply some form of disclosure limitation to the verification measures, ideally one that enables the data steward to bound the information leakage.  This is precisely what differential privacy promises, which motivates us to develop the new verification measures presented in Section \ref{verification}.



Undoubtedly, some analyses will not be adequately preserved by the synthetic data. The verification server will help
users learn this, thereby reducing the chances of false findings based on the synthetic data.  These users may desire access to
the confidential data, which motivates the third prong of the integrated data access system: provide
remote access to  confidential data to approved users via virtual machines on a protected research
  data network (PRDN).   Variants of PRDNs are in use by many organizations, including national statistical agencies, universities, and the National Opinion Research Center.  Thus, we do not describe how to set up the architecture for a PRDN in this article, although we use one at Duke University (\url{https://arxiv.org/abs/1710.03317}) to validate results in the OPM application.




Integrating all three ideas in a single system creates synergies. Data stewards can establish policies with low barriers for access to the synthetic data, for example, by allowing users to access the synthetic data without completing an extensive approval process.   
 Users can start with the synthetic data to investigate distributions and relationships, determine what questions might be answerable with the data (e.g., are there enough cases of interest to support accurate modeling?), examine the need for transformations and recoded variables, and develop appropriate code.  
The verification server can enable users to know when to trust and act on their results, 
and when perhaps not to do so.  Even users who are not satisfied with the quality of the results can benefit from starting with the synthetic data.
Storage and processing of large-scale data are costly to data stewards, who likely will pass some costs to users.  Users who have an
informed analysis plan---for example, they know the approximate
marginal distributions of the data and  have a sense of the data structure---can improve their efficiency when using the PRDN, thereby saving their own time and money.  By performing their data explorations outside the PRDN, these analysts will use up fewer cycles on the protected systems and open opportunities for more efficient use of those systems. 


\section{Description of  Data}\label{background}


\subsection{Overview of the confidential data}\label{opmdata}

The Office of Personnel Management  maintains the personnel
records for all civil servants in the United States. 
 We work with a subset of the data from the  OPM's Central Personnel Data File (CPDF) and Enterprise Human Resources Integration system (EHRI), which we jointly refer to as the Status File (SF).  
The SF we use is a snapshot of the civil service on every September 30 (the end of the fiscal year), comprising approximately two million employees per year from 1988 to 2011.  For each employee, the file includes annual data on  characteristics like age, agency, education level, pay grade,
occupation, supervisory status,  entry and departure, and other
background characteristics.  The data are longitudinally linked.
We exclude employees from the armed services, the Department of Defense, the U.\ S.\ Postal Service, and individuals who work in classified roles, sensitive agencies, and sensitive occupations as defined by OPM. The final analysis file includes personnel records from 3,511,824 employees.








The OPM data are valuable because they allow researchers to investigate many key questions in the study
of human capital in large organizations and government organizations
in particular.  For example, 
researchers can use the OPM data to 
examine government agencies' ability to recruit high quality individual
talent, to develop their employees' expertise within those agencies,
and to retain the best and brightest in government service \citep[e.g.,][]{lewis95, bolton2}.  
These are important and complicated challenges; public agencies must
cope with episodic turnover of political appointees, limited ability
to adjust worker compensation in response to outside market pressures,
difficulty in performance measurement due to the nature of
governmental tasks, and constraints on frictionless alterations to the
government workforce because of employment terms for civil
servants \citep{borjas80, bolton1}. Ultimately, research with these data 
can shed light on the relative costs and benefits of  human capital management strategies.   
The OPM data also are used by the government as a major input for wage comparisons, tracking personnel, determining workforce diversity, and helping to construct personnel policies for the over two million civil servants.

As of this writing, the OPM has made several derivatives of the SF data available to the public, although not the full SF data that we work with. In particular, the OPM releases summary data online  as ``data cubes.'' These comprise quarterly tabular summaries of a limited set of variables without employee indicators that would allow for over-time comparisons. Pursuant to the Freedom of Information Act (FOIA), the OPM provides non-anonymized data for six variables---name, agency, location, position, grade, and salary---for each civil service employee in a non-sensitive position to anyone who requests them. Some proprietary websites, such as FedSmith, also make some or all of these limited  data available to the public. The news agency BuzzFeed used a FOIA request to obtain and subsequently release a subset of the OPM data. Unlike our SF data, the data released by BuzzFeed do not include race or gender, which typically are not included by the OPM in FOIA requests, nor 14 other variables that we synthesize.  

%


We obtained the SF data through an agreement with OPM, with the understanding that we would not reveal personally identifiable information when using the data, especially information that OPM deems sensitive like race and gender.
The OPM data files are currently housed in a PRDN at Duke University with strict access
and confidentiality standards under a Duke University IRB-approved protocol.
We did extensive work cleaning and preparing these data for research purposes, in accordance with the OPM's ``Guide to Data Standards (Part A).'' 
The supplemental material in \citet{barrientos;etal;2017a} includes more details on our data preparation processes. 

The availability of the limited data, as well as now the BuzzFeed data and possibly future releases from FOIA requests, pose an obvious 
disclosure risk problem. If the SF data were
``anonymized'' by standard techniques employed by government agencies---e.g., stripping names coupled with aggregating or perturbing a small fraction of the data values---an ill-intentioned user 
might be able to reverse engineer a large percentage of the OPM database by
matching the six or more known fields to the same fields in the anonymized
data.  They subsequently could retrieve the private and confidential data of a
large percentage of the federal personnel.

\subsection{Overview of synthetic data creation}\label{syntheticdata}

%



To reduce these disclosure risks, the OPM could release a fully synthetic version of the SF. 
In this section, we provide an overview of our process for generating the synthetic SF data.
We focus on the methods for generating synthetic careers, races, and wages; these variables 
are central to our illustrative verification analysis of wage differentials. The synthesis models for other variables are described 
in \citet{barrientos;etal;2017a}. As of this writing,  the OPM has not yet determined whether or not  to make the synthetic data available to a broader set of researchers.

The SF data are complicated, making the task of generating useful synthetic data challenging.
The employees are measured on $29$ variables over the course of $24$ years.  
They work across 607 agencies, some of which have only a handful of employees and 
some of which have thousands of employees.   
The variables are mostly nominal with levels ranging from 2 (sex) to 803 (occupation), and also include a small number of numerical variables.  For any employee, most variables can change annually, although a few demographic variables remain constant 
or change deterministically (age). 
Many pairs of variables have theoretically impossible combinations, such as certain occupations being restricted to certain education and degree types.
Some variables should be non-decreasing over time, such as months of military service and educational levels, although the confidential data have records that violate those restrictions, presumably due to reporting errors.

To make the synthetic data, we construct a joint distribution using sequential conditional modeling, as done in \citet{lbdisr}.  This allows us to develop models targeted to different types of variables, to conveniently incorporate logical relationships among the variables, and to develop efficient, parallelizable code.  We order the variables from the first to last to be synthesized.  
Let ${\bf V}_{ij}$ be the value of the $j$-th ordered variable for the $i$-th employee, where $j=1, \dots, 29$.  We seek the joint distribution,  
\begin{equation}
p({\bf V}_{i1},\ldots,{\bf V}_{i29}) = p_1({\bf V}_{i1}) \times p_2({\bf V}_{i2}|{\bf V}_{i1}) \times  \ldots \times p_{29}({\bf V}_{i29}|{\bf V}_{i1},\ldots,{\bf V}_{i28}),
\end{equation}
where 
each $p_j$ denotes the conditional distribution of ${\bf V}_j$ given ${\bf V}_1,\ldots,{\bf V}_{j-1}$. 
We let ${\bf V}_1$ correspond to the sequence of agencies where the employee has worked, which defines the employee's career. 
Nearly all other variables depend on when and where the employee works, so modeling this variable first 
facilitates the synthesis process.
We let $({\bf V}_2, \dots, {\bf V}_7)$ be, in order, gender, race, educational level, age in years,  years since the employee 
earned the degree mentioned in educational level, and an indicator for ever having served in the military.
These demographic variables are, for the most part, straightforward to model because either (1) they remain constant  
across time or change in a deterministic manner after the initial year, or (2) change with only low 
probabilities. We  let $({\bf V}_8,\ldots,{\bf V}_{29})$ include the remaining variables, which  depend on the characteristics of the employee's job 
that year. Examples of these variables include occupation, part-time or full-time status, grade and step classification, supervisory status, and pay. A full list of variables is in \citet{barrientos;etal;2017a}. 

For ${\bf V}_j$ that can change annually, where $j> 1$, we generally apply lag-one modeling strategies to simplify computation.  Specifically, 
let $V_{ijt}$ be the $j$-th variable at year $t$ for the $i$-th employee. Let $t_{i1} < \ldots < t_{in_i}$ be the $n_i$ years when employee $i$ has values (is working),  
and set ${\bf V}_{ij} = (V_{ijt_{i1}}, \ldots, V_{ijt_{in_i}})$. For longitudinal ${\bf V}_j$, we use the 
conditional representation, 
\begin{eqnarray}\label{longit_model1}
p_{j}\left({\bf V}_{ij}\left.|{\bf V}_{i1},\ldots,{\bf V}_{i(j-1)}\right.\right) =  \prod_{l=2}^{n_i} p_{jt_{il}}\left(V_{ijt_{il}} \left|{\bf V}_{i1},\ldots,{\bf V}_{i(j-1)}, V_{ijt_{i1}} ,\ldots, V_{ijt_{il-1}}\right.\right), 
\end{eqnarray}
where $p_{jt_{il}}$ denotes the distribution of $V_{ijt_{il}}$ conditioned on the previous $j-1$ variables and the values of ${\bf V}_{ij}$ up to time $t_{il-1}$. 
We assume that
\begin{eqnarray}\label{longit_model2}
& & \hspace{-15mm}
p_{jt_{il}}\left(V_{ijt_{il}} \left| {\bf V}_{i1},\ldots,{\bf V}_{i(j-1)}, V_{ijt_{i1}}, \ldots, V_{ijt_{il-1}}\right.\right) =  
p_{jt_{il}}\left(V_{ijt_{il}} \left| V_{i1t_{il}},\ldots, V_{i (j-1)t_{il}}, V_{ijt_{il-1}}\right.\right).
\end{eqnarray}
Thus, the conditional distribution of $V_{ijt_{il}}$ depends only on current values of ${\bf V}_{ij'}$, $1<j'<j-1$, and the nearest 
past value of ${\bf V}_{ij}$. 

For purposes of this article, we create and analyze one synthetic dataset.  If desired by the OPM, we could create and release multiple synthetic datasets by repeating the data generation process.  An advantage of releasing multiple synthetic datasets is that users can propagate uncertainty from the synthesis process through their inferences using simple combining rules \citep{raghu:rubin:2001, reiter:raghu:07, drechslerbook}.  

\subsubsection{Modeling strategy for employees' careers (${\bf V}_1$)}
We define an employee's career as the sequence of agencies where the employee has worked throughout the $24$ years. 
Since most employees have not worked in all $24$ years, 
 we create an additional ``agency'' corresponding to the status of not working. 
With this additional level, all employees' sequences have length 24.

To model these sequences, we create three additional variables. Let 
$G$ be the number of agencies where the employee has worked over the 24 years.  Let 
${\bf Z}$ be the list of years in which the employee moved to a new agency, including a change in working status. 
Let ${\bf W}$ be the ordered sequence of unique agencies where the employee has worked.
The values of $(G, {\bf Z}, {\bf W})$ completely describe the entire career of any employee, as illustrated in 
Table \ref{sec2:tab1}.  


\begin{table}[t]
\begin{center}
\begin{tabular}{c|llllllllll|ccc}
  Employee &  \multicolumn{10}{|c|}{Employee's career} & $G$ & ${\bf Z}$ & ${\bf W}$ \\\hline
  $e_1$ &  0 & 0 & A & A & 0 & 0 & C & C & C & C &   3 & (3,5,7) & (0,A,0,C)\\
  $e_2$ &  0 & 0 & 0 & 0 & 0 & 0 & 0 & 0 & 0 & B & 2 & 10 & (0,B)\\
  $e_3$ &  A & 0 & B & C & C & A & A & A & 0 & 0 & 4 & (2,3,4,6,9) & (A,0,B,C,A,0)\\
\end{tabular}
\caption{Illustration of how to define $(G, {\bf Z}, {\bf W})$ using three hypothetical employees and 10 years. 
Each column in the employee's career represents a year; a 0 means the employee did not work that year; and, 
A, B, and C  represent three different agencies.  For example, employee $e_1$ did not work in years 1 and 2, worked in  agency A for two years, stopped 
working in years 5 and 6, and worked in agency C during the last four years. \label{sec2:tab1}}
\end{center}
\end{table}



Defining a model for employees' careers is equivalent to defining a model for $(G, {\bf Z}, {\bf W})$, which we do sequentially. 
For $G$, we use a discrete distribution on $\{1, \dots, 24\}$ with probabilities equal to the observed frequencies of each value, from which we randomly generate the values of 
$G$ associated with the synthetic employees. 
For ${\bf Z}|G$, we create a one-to-one function $T_G$ to map ${\bf Z}$ into a space of permutations dependent on $G$. 
We model $T_G({\bf Z})|G$ using a latent model defined on the simplex space. The latent model is defined using mixtures of Dirichlet distributions. 
This model allows us to borrow information across different agency patterns (given $G$) and, therefore, to give positive probability to unobserved values of ${\bf Z}|G$. 
Since we use a one-to-one mapping, the model for $T_G({\bf Z})|G$ can be easily used to generate values from ${\bf Z}|G$. 
Finally, we model ${\bf W}|{\bf Z},G$ using a Markov chain of order one. A formal description of the three sub-models is in \citet{barrientos;etal;2017a}.
While targeted at modeling careers, this model can be applied more generally for other sequences of categorical variables.

\subsubsection{General strategy for race (${\bf V}_3$)}

Almost all employees report the same value of race in all 24 years. However, 2.7\% of employees report different values across the years, usually changing  values only once or twice. It is possible that these represent clerical errors in the data, but it is also the case that OPM changed the ways that they handled the collection of race data during the period covered by our dataset. In particular, different categories were available for employee selection at different points from 1988 to 2011, and eventually OPM allowed individuals to select more than one race. There is no way for us to distinguish clerical errors from instances in which employees changed their race identification for personal reasons or because of how the data were collected.

 Rather than model longitudinal changes in race across time for all employees, which easily could result in far more switching than observed in the data, we instead create an auxiliary binary variable that indicates whether the values of race remain the same across all years or not. Using the confidential data, we estimate a model for this binary outcome with classification and regression trees (CART), conditioning on sex  (${\bf V}_2$) and predictors derived from the employee's career (${\bf V}_1$). CART models are useful as synthetic data engines, as they can reflect important features of conditional distributions automatically and flexibly.  Following the approach in \citet{reitercart}, we run the synthetic values of (${\bf V}_1, {\bf V}_2$) down the fitted tree to generate synthetic values of the binary variables for each synthetic employee.  For synthetic employees whose generated binary variable indicates that their race value does not change, we predict their race using a CART-based model estimated from the confidential data. This model uses the first observed race as the outcome variable and $({\bf V}_1, {\bf V}_2)$ as predictors.  Finally, for synthetic employees whose binary variable indicates that their race values change across time, we model the race at each year using \eqref{longit_model1} and \eqref{longit_model2}, using CART for each $p_{3,t}$ where $t=1,\ldots,24$.

%

\subsubsection{General strategy for wages (${\bf V}_{27}$)}  

Federal employees' basic pay (salary before any location adjustments) is set by tables known as   
pay plans. For example, most government employees in white collar occupations fall under the General Schedule pay 
plan. For most pay plans, basic pay is a deterministic function of 
a combination of variables, usually including the employee's grade and step. Thus, in theory, you can find any federal employee's pay 
by locating their grade and step  on their pay plan table.    
However, in the SF, some employees' basic pay is not consistent with their pay plan, grade, and step; when this happens, usually the pay coincides with a value in the pay table associated 
with a neighboring step. 

To capture these features, we synthesize pay plan, grade, and step before basic pay, 
thereby allowing us to ``look up'' the pay for the synthetic employees.  We   
model basic pay using  
\eqref{longit_model1} and \eqref{longit_model2} with a CART synthesizer, assuming that basic pay is a 
nominal variable. This essentially is equivalent to sampling from the values of 
basic pay reported in grade and step for a given pay plan, but also allowing for other variables on the
file to explain deviations from the pay plan. 





\subsubsection{Evaluations of analytic validity} \label{utilityevals}

We evaluate the analytic validity of the synthetic OPM data using the general approach for synthetic data products currently published by federal statistical agencies. First, we consult subject matter experts and staff at the OPM to determine a set of analyses that is representative of likely uses of the synthetic data.  The subject matter experts suggested estimands involving race and gender, which are not available for individual employees on public use files, as well as analyses involving longitudinal trends. The staff at the OPM suggested estimands that can be obtained from the data cubes, as they believe users will have more faith in the synthetic data if those are faithfully reproduced.  Second, after selecting the analyses, we run them on the synthetic and confidential data within the PRDN.  We compare the results, identifying features of the joint distribution that are modeled accurately and not as accurately as desired.  This process is iterative, in that we use the results to refine and improve the synthesis models.  

With a dataset of this complexity and dimensionality, the number of potentially interesting analytical validity checks is enormous.  We summarize an extensive, but necessarily partial, set of results from these checks in \citet{barrientos;etal;2017b} and at  \url{https://github.com/DukeSynthProj/DukeSyntheticDataResources}. In these results we include analyses of pay differentials by race for sub-populations of the OPM file, including specific agencies and occupations.  Of course, with a dataset of this complexity and dimensionality, it is practically impossible to preserve all relationships when synthesizing.
This underscores the benefits of verification servers for synthetic data products.

\subsubsection{Evaluations of disclosure risks} \label{riskevals}

For this synthetic OPM file, we do not evaluate the risks that intruders can use the synthetic data to learn whether or not particular individuals are in the confidential OPM database.  The confidential OPM database is a census, so that all federal employees (in the agencies we include) are known to be in the database.  Further, as mentioned in Section \ref{opmdata}, various non-anonymized data products  derived from the confidential OPM data are publicly available. Hence, intruders seeking to learn whether or not certain individuals are in the OPM database have far more informative---indeed, perfectly informative---sources to use than the synthetic OPM data.  Instead, we focus on the risk that intruders can use the synthetic OPM data to infer values of individual employees' sensitive variables that are not in the public data products, like gender and race.  

To assess these inferential disclosure risks, we compute probabilities that intruders correctly guess individuals' sensitive values given the synthetic data.  Here, we describe an approach for assessing inferential disclosure risks for employees' most recently reported races. 
  Broadly, we identify combinations of variables  known to be publicly available, namely the six fields released as non-anonymized data by OPM and the 14 variables in the BuzzFeed data, that are present in both the synthetic and confidential OPM files.  For each of these combinations, we compute the percentage of times each race category appears in the synthetic data.  We record the numbers of individuals for whom the percentage corresponding to their actual race exceeds a threshold deemed too high risk (thresholds are determined by policies of the data steward).  Details of the methodology appear in \citet{barrientos;etal;2017a}.

Revealing the thresholds and the exact numbers of individuals exceeding them could result in increased disclosure risks for some records.  Hence, as typical when reporting disclosure risks in applications with genuine sensitive data \citep[e.g., ][]{holan10}, we do not publish precise estimates of the probabilities.  We can say, however, that less than 15\% of employees' true races can be guessed correctly by using this attack strategy.  Whether or not this is an acceptable risk is, of course, a matter for the data steward (OPM) to decide. 

When considering such probabilities, it is also important for data stewards to compare against relevant baseline risks.  For example, suppose that anyone can learn the marginal distribution of race for particular agencies in particular years from public sources.  Suppose further that this public information reveals that all employees in a certain agency report the same race in one year (which happens for some small agencies).  In this scenario, synthetic data with this same marginal distribution do not reveal any new information about these employees' races. Arguably, this should not be counted as a disclosure risk attributable to the synthetic data.  In fact, 
the data cubes published by OPM include tables of marginal distributions of race by agency, as well as margins of race by other variables.  Hence, we can treat the marginal distributions of race from the cubes as baseline risks.  
Doing so, we find that more than half of the races correctly estimated in the synthetic data also can be determined using just the marginal information from the data cubes.  
We also find that, across all employees in one particular year, the probabilities for the synthetic OPM data are lower than those for the data cubes 
for about 50\% of employees, and higher for about 15\% of employees. Again, whether or not this is an acceptable incremental disclosure risk is a policy decision for the data steward rather than the statisticians who assess the risk.


Assessing inferential disclosure risks for high-dimensional, fully synthetic data is a computationally intensive and challenging task, and is the subject of ongoing research. For additional discussion, see \citet{reiter:2002a}, \citet{abowdvilh}, \citet{drechslerbook}, \citet{wangreiter}, \citet{synriskzhang}, and \citet{mcclurereiter16}, as well as the applications cited in Section \ref{SynthVerif}.  Section \ref{Conclusion} includes additional discussion of disclosure risks in the integrated system.







\section{Verification Measures}\label{verification}


%
%
%

The quality of inferences from synthetic data depend entirely on the quality of the models used to generate the data, 
as the synthetic data only can reflect distributional assumptions in the synthesis models \citep{reiter:2002a}.
Analysts of the synthetic data need some way to
 assess the accuracy of their particular inferences based on the synthetic data.
Verification servers provide an automated means to  provide  such feedback.


In designing a verification server, the agency must account for a crucial fact: verification
measures leak information about the  confidential data  \citep{vs08, mcclurereiter}. Clever data snoopers could submit queries that, perhaps in combination with other information, allow 
them to estimate confidential values too accurately.  To reduce  these risks,
one approach is to require verification measures to satisfy $\epsilon$-differential privacy ($\epsilon$-DP), which we now explain briefly. 

Let $\mathcal{A}$ be an algorithm that takes as input a database $\mathbf{D}$ and outputs some quantity $o$, i.e., $\mathcal{A}(\mathbf{D}) = o$.  In our context, these outputs are used to form verification measures.  Define neighboring databases, $\mathbf{D}$ and $\mathbf{D}'$, as databases of the same size that differ in one row and are identical for all other rows.


\begin{definition}[$\epsilon$-differential privacy.]
An algorithm $\mathcal{A}$ satisfies $\epsilon$-differential privacy if for any pair of neighboring databases $(\mathbf{D}, \mathbf{D}')$, and any output $o \in range(\mathcal{A})$,
the $Pr(\mathcal{A}(\mathbf{D}) = o) \leq \exp(\epsilon) Pr(\mathcal{A}(\mathbf{D}') = o).$
\end{definition}
Intuitively, $\mathcal{A}$ satisfies $\epsilon$-DP when the distributions of its 
outputs are similar for any two neighboring databases, where similarity is defined by the factor $\exp(\epsilon)$.   The $\epsilon$, also known as the privacy budget, controls the degree of the privacy offered by $\mathcal{A}$, with lower values implying greater privacy guarantees. $\epsilon$-DP is a strong criterion, since
even an intruder who has access to all of $\mathbf{D}$ except any one row learns little from $\mathcal{A}(\mathbf{D})$  about the values in that unknown row when $\epsilon$ is small. 


Differential privacy has three other properties that are appealing for verification measures.  
Let $\mathcal{A}_1(\cdot)$ and $\mathcal{A}_2(\cdot)$ be $\epsilon_1$-DP and $\epsilon_2$-DP algorithms.  First, 
for any database $\mathbf{D}$, releasing the outputs of both $\mathcal{A}_1(\mathbf{D})$ and $\mathcal{A}_2(\mathbf{D})$ ensures $(\epsilon_1 + \epsilon_2)$-DP.  Thus, we can track the total privacy leakage from releasing verification measures.  Second, releasing the outputs of both $\mathcal{A}_1(\mathbf{D}_1)$ and $\mathcal{A}_2(\mathbf{D}_2)$, where $\mathbf{D}_1 \cap \mathbf{D}_2 = \emptyset$, satisfies $\max \{\epsilon_1, \epsilon_2\}$-DP.  Third, for any algorithm $\mathcal{A}_3(\cdot)$, releasing $\mathcal{A}_3(\mathcal{A}_1(\mathbf{D}))$ for any $\mathbf{D}$ still ensures $\epsilon_1$-DP.  Thus, post-processing the output of $\epsilon$-DP algorithms does not incur extra loss of privacy.


One method for ensuring $\epsilon$-DP, which we utilize for $\epsilon$-DP verification measures, is the Laplace Mechanism \citep{dwork:06}.
For any function $f : \mathbf{D} \rightarrow \mathbb{R}^d$, let $\Delta(f) = \max_{(\mathbf{D_1},\mathbf{D_2})} ||f(\mathbf{D}_1) - f(\mathbf{D}_2)||_{1}$, where $(\mathbf{D}_1, \mathbf{D}_2)$ are 
neighboring databases.  This quantity, known as the global sensitivity of $f$, is 
the maximum $L_1$ distance of the outputs of the function $f$ between any two neighboring databases.
%
The Laplace Mechanism is $\mathbf{LM}(\mathbf{D}) = f(\mathbf{D}) + \eta$, where $\eta$ is a $d \times 1$ vector of 
independent draws from a Laplace distribution with density $p(x \mid \lambda) = (1/(2\lambda)) \exp(-|x| / \lambda)$, where $\lambda = \Delta(f) / \epsilon$.


We now present verification measures that satisfy $\epsilon$-DP and help analysts assess the importance of regression coefficients. 
We derive the measures for linear regression, as we use these models in the analysis of wage differentials by race. 
To fix notation for describing the measures, let $\bD$ include all $n$ individuals in  
the subset of the confidential data that is of interest for analysis. For any individual $i$ belonging to $\bD$, let $y_i \in \mathbb{R}$ be the response variable and $x_i = (1,x_{i,1},\ldots,x_{i,p})^T \in \mathbb{R}^{p+1}$ 
be a set of predictors, where both are transformed as desired for regression modeling.  Hence, $\bD=\{(x_i, y_i)\}_{i=1}^n$, where
all values are from the confidential data. Let $E(y_i \mid x_i) =  \beta^T x_i$, 
where $\boldsymbol{\beta} = (\beta_0,\ldots,\beta_p)^T \in \mathbb{R}^{p+1}$.

\subsection{Measures for importance of regression coefficients}\label{sec:ver1}


In many contexts, analysts are interested in whether or not the value of some $\beta_j$ 
exceeds some threshold, say $\gamma_0$.  For example, users of the SF data (economists, policy makers, lawyers)
might consider a value of $\beta_j$ corresponding to 1\% or larger differential in average pay  
to be practically significant evidence of wage discrimination.   But, 
they might be less concerned when $\beta_j$
corresponds to a differential less than 1\%.   Without loss of generality, assume that we want to determine if 
some $\beta_j < \gamma_0$. Corresponding to this decision, we define the parameter   
$\theta_0 = \mathbb{I}_{(-\infty, \gamma_0]}(\beta_j)$, 
where $\mathbb{I}_{(-\infty, \gamma_0]}(\beta_j)$ is an indicator function that equals one when $\beta_j \in (-\infty, \gamma_0]$ and equals zero otherwise.
We note that the measure can be used for any interval, for example, of the form $[l,u]$ or $(u, \infty]$.  



For many regression analyses of the SF data,  the confidence intervals for the relevant $\beta_j$  are narrow due to large sample sizes. 
For these $\beta_j$ and most $\gamma_0$, the maximum likelihood estimate (MLE) of $\beta_j$ effectively tells the analyst whether 
$\theta_0 = 1$ or $\theta_0 = 0$. To formalize this notion, let $\hat{\beta}_j^N$ be the MLE of $\beta_j$ based on a sample with $N$ individuals (where $N$  stands for a generic sample size).  We approximate $\theta_0$ by using the pseudo-parameter, 
$$
\theta_N = 
\left\{
\begin{array}{ll}
1 & \mbox{if } P[\hat{\beta}_j^N \leq \gamma_0 ] \geq \gamma_1, \\
0 & \mbox{if } P[\hat{\beta}_j^N \leq \gamma_0 ] < \gamma_1.
\end{array}
\right.
$$   
Here, $\gamma_1 \in (0,1)$ reflects the degree of certainty required by the user before she decides there is enough evidence 
to conclude that $\theta_0 = 1$. 
When $\hat{\beta}_j^N$ is a consistent estimator of $\beta_j$, we can guarantee that  
$\lim_{N \rightarrow \infty} \theta_N = \theta_0$.









Unfortunately, we cannot release $\hat{\beta}^N_j$, nor other deterministic functions of $\bD$, directly and satisfy $\epsilon$-DP. 
Instead, we release a noisy version of the key quantity in $\theta_N$, namely $r = P(\hat{\beta}_j^N \leq \gamma_0 )$.  
We do so using the sub-sample and aggregate method \citep{nissim;raskhodnikova;smith;2007}. 
We randomly split $\bD$ into $M$ disjoint subsets, $\bD_1,\ldots,\bD_M$, of size $N$ (with inconsequential differences when $N=n/M$ is not an integer), where 
$M$ is selected by the user.  We discuss the choice of $M$ in Section \ref{Conclusion}. In each $\bD_l$, where $l=1, \dots, M$, we compute  the MLE $b_{jl}$ of $\beta_j$. The $(b_{j1},\ldots,b_{jM})$  can be treated as $M$ independent draws from the 
distribution of $\hat{\beta}_j^N$, where $N = n/M$.  Let $W_{l} = \mathbb{I}_{(-\infty, \gamma_0]}(b_{jl})$. Each 
$W_{l}$ is an independent, Bernoulli distributed random variable with parameter $r$. 
Thus, inferences for $r$ can be made based on $S = \sum_{l=1}^M W_l$. 
However, we cannot release $S$ directly and satisfy $\epsilon$-DP; instead, we generate a noisy version of $S$ using the Laplace Mechanism with $\lambda = 1/\epsilon$, 
resulting  in $S^R = S + \eta$.
The global sensitivity equals $1$, since at most one of the partitions can switch from zero to one (or vice versa).


The noisy $S^R$ satisfies $\epsilon$-DP; however, interpreting it directly can be tricky.  First, $S^R$ is not guaranteed to lie in $(0,M)$ nor even to be an integer.  Second, 
alone $S^R$  does not provide estimates of uncertainty about $r$.  We therefore use a post-processing step---which has no
bearing on the privacy properties of $S^R$---to improve interpretation.  We find the posterior distribution of $r$ 
conditional on $S^R$ and using the noise distribution, which is publicly known. Using simple Markov chain Monte Carlo techniques, we estimate the model, 
\begin{equation}\label{bayes_model}
S^R|S  \sim {\rm Laplace}(S, 1/\epsilon),\,\,\,\,\  S \mid r    \sim {\rm Binomial}(M,r), \,\,\,\,\,  r      \sim \rm{Beta}(1,1).
\end{equation} 
Here, we treat $S$ as an unobserved random variable and average over it.  Hence, the computations never touch the 
confidential data other than through the differentially private $S^R$.  
  
The verification server reports back the posterior distribution of $r$ to the analyst, who can approximate $\theta_N$ 
for any specified $\gamma_1$ simply by finding the amount of posterior mass below $\gamma_1$.  Alternatively, 
analysts can interpret the posterior distribution for $r$  as a crude approximation to the Bayesian posterior probability, $\pi\left(\beta_j \leq \gamma_0 |S^R\right)$.  
For instance, if the posterior mode for $r$ equals $0.87$, we could say that 
the posterior probability that $\beta_j < \gamma_0$ is approximately equal to $0.87$.  
We caution that this latter interpretation may not be sensible for small sample sizes.

In the verification of the OPM regression analyses, we use  measures that compare differences between regression coefficients estimated with the confidential data and user-specified thresholds.  The substantive experts on our team felt that this was most appropriate for the analysis of wage gaps.  However, the measures can be used to compare synthetic and confidential data regression coefficients.  Rather than setting thresholds based on scientific or policy considerations, the user can set thresholds based on the differences they are willing to  tolerate between the synthetic and confidential data coefficients.  For example, suppose the synthetic data coefficient of interest is -.021. Suppose that the user considers the synthetic data estimate sufficiently accurate when the confidential data coefficient is within $\pm$50\% of the synthetic data coefficient (or some other user-defined tolerance interval).  The user can set the threshold at $(l, u) =  (-.031, -.010)$.  A reported posterior mode of $r$ near 1 suggests that the difference between synthetic data and observed data coefficients is within the tolerance bound ($\pm$ 50\%), whereas a value near zero suggests otherwise.

For any particular analysis, the usefulness of these verification measures depends on the sample size, the number of partitions, and the value of $\epsilon$ allowed by the data steward. \citet{barrientos;etal;2017a} includes analyses of the OPM data that illustrate the performance of the verification measures for different values of these features. 
We discuss some of these findings in Section \ref{Conclusion}.

\subsection{Measures for longitudinal trends in regression coefficients}\label{sec:ver2}



With longitudinal data, analysts often are interested in how the value of some $\beta_j$ 
changes over time. For example, in the SF analysis, we want to know whether the racial wage gap is closing or growing as the years advance. 
Suppose for a moment that we knew the values of $\beta_j$ for all years.  One simple way to characterize the trend  in $\beta_j$ over time is to break the data into $K$ consecutive periods and, in each time period, find the OLS line  
predicting $\beta_j$ from year.  The slopes of these lines pasted together represent a piece-wise approximation to the trend. Of course, 
we do not know the values of $\beta_j$; we must use $\bD$  to learn about these slopes. 

We use this idea to construct a verification measure for longitudinal trends in regression coefficients.  Specifically, the analyst begins by selecting $K$ periods of interest.  In each period, the analyst posits some interval for the slope and 
requests an $\epsilon$-DP verification of whether the values of $\beta_j$ are consistent with that posited interval.  For example,  
the analyst might split $\bD$ in $K=2$ consecutive intervals, and posit that the slope of $\beta_j$ over the years is negative in the first period and positive in the second period. In the wage gap 
analysis, this would correspond to a growing wage gap in the first period, followed by a shrinking wage gap in the second period. The analyst can use the synthetic data to identify the periods of interest and 
set the intervals for the slopes, as we illustrate in Section \ref{wage:annual}. Effectively, this evaluates whether the trends in $\beta_j$ estimated with the confidential data match the trends estimated with the synthetic data.



Formally, suppose that $\bD$ can be divided into nonempty subsets, $\{\bD^t\}_{t\in\mathcal{T}}$, where $\bD^t$ denotes 
all the data points in $\bD$ at year $t$, and $\mathcal{T}$ is some period of years under study.
Further, suppose that for every $(y_{it},x_{it}) \in \bD^t$, 
$E(y_{it}|x_{it})= \boldsymbol{\beta}_t^T x_{it}$, where $\boldsymbol{\beta}_{t}=(\beta_{0t},\ldots,\beta_{pt})^T$ is the vector of coefficients at time $t$. 
Let $\mathcal{T}_k \subset \mathcal{T}$ be a subset of years.  The analyst seeks to learn the overall trend in the values of $\beta_{jt}$, where $t \in \mathcal{T}_k$, during that time. 
To characterize this trend, let   
 $m\left(\{(t,\beta_{jt})\}_{t \in \mathcal{T}_k}\right)$ be a real-valued 
function that returns the slope of the OLS line passing through the points $\{(t,\beta_{jt})\}_{t \in \mathcal{T}_k}$.  
The analyst might be interested in, 
for example, whether $m\left(\{(t,\beta_{jt})\}_{t \in \mathcal{T}_k}\right) < 0$ indicating a decreasing trend, 
$m\left(\{(t,\beta_{jt})\}_{t \in \mathcal{T}_k}\right) > 0$ indicating an increasing trend, 
or more generally, $m\left(\{(t,\beta_{jt})\}_{t \in \mathcal{T}_k}\right) \in C_k$ for some interval $C_k$, e.g., $C_k$ is tight around zero for a flat trend.
Hence, for any interval $C_k$, the analyst seeks to learn  
$\theta_0 = \mathbb{I}_{C_k}\left(m\left(\{(t,\beta_{jt})\}_{t \in \mathcal{T}_k}\right)\right)$, 
where  $\mathbb{I}_{C_k}\left(m\left(\{(t,\beta_{jt})\}_{t \in \mathcal{T}_k}\right)\right)$ is an indicator function that equals one when $m\left(\{(t,\beta_{jt})\}_{t \in \mathcal{T}}\right) \in C_k$ and equals 
 zero otherwise. 

Because $\theta_0$ is a binary parameter, we can use the methods in Section \ref{sec:ver1} to release an $\epsilon$-DP version of it.  Here we outline the procedure; formal details are in \citet{barrientos;etal;2017a}.
We split $\mathbf{D}$ into $M$ partitions of employees.  In each $\mathbf{D}^t_l$, we compute the MLE $b_{jtl}$ of $\beta_{jt}$. 
 We let $W_{l}= \mathbb{I}_{C_k}(m\left(\{(t,b_{jtl})\}_{t\in\mathcal{T}_k}\right)$ and $S=\sum_{l=1}^M W_l$. Following the logic of Section \ref{sec:ver1}, we use \eqref{bayes_model} to get posterior inferences for $r = P[m(\{(t,\hat{\beta}_{jt}^{N_t})\}_{t \in \mathcal{T}_k}) \in C_k]$, where $\hat{\beta}_{jt}^{N_t}$ is the MLE of $\beta_{jt}$ based on a sample with $N_t$ individuals.



When trends over the entire $\mathcal{T}$ are of interest, analysts can partition $\mathcal{T}$ into $K$ consecutive 
periods, $\mathcal{T}_k=\{t_{k-1},t_{k-1}+1,\ldots,t_k\}$, where $k=1,\ldots,K$ and $t_0 < t_1 < \ldots
 < t_K$.  For a given set of intervals $\{C_k\}_{k=1}^K$, the analyst can do the verification separately for each interval, and interpret the set of results.  Alternatively, the analyst can perform a single verification across all intervals, setting the parameter of interest to 
 $\theta_0 = \prod_{k=1}^K \mathbb{I}_{C_k}\left(m\left(\{(t,\beta_{jt})\}_{t \in \mathcal{T}_k}\right)\right)$. 
Here,  $\theta_0$ equals one when $m\left(\{(t,\beta_{jt})\}_{t \in \mathcal{T}}\right) \in C_k$ for every $k=1,\ldots,K$, and equals 
 zero otherwise.  For example, to examine whether the trend of $\beta_{jt}$ is decreasing during 
the first 9 years and is increasing during the last 15 years, the analyst would set $C_1=(-\infty,0)$, $C_2=(0,\infty)$, $\mathcal{T}_1=\{1,\ldots,9\}$, and $\mathcal{T}_2=\{10,\ldots,24\}$.  If the mode of the posterior probability for $r$ equals $0.93$, we say that the posterior probability that $\beta_{jt}$ decreases during the first 9 years and increases over the last 15 years approximately equals $0.93$.

When setting $C_k$ to $(-\infty, 0)$ or $(0, \infty)$, i.e., simply estimating whether the slope of the values of $\beta_j$ over $\mathcal{T}_k$ is negative or positive,
the posterior modes have predictable behavior. Values are close to one when the true slope has the sign implied by $C_k$ and is far from zero; values are 
close to zero when the true slope has opposite sign and is far from zero; and, values are close to .5 when the true slope itself is close to zero.  This last feature arises when  the slopes in the partitions bounce randomly around zero.


The analyst who requests a single verification for $\mathcal{T}$ spends only $\epsilon$ of the privacy budget.  However, this analyst only can tell if the whole  
trend over $\mathcal{T}$ in the confidential data matches that in the synthetic data. In contrast, the analyst who requests $K$ verifications, one for each $\mathcal{T}_k$, spends $K\epsilon$ of the budget.  But, this analyst gets finer details of the trend.  For this reason, when the privacy budget allows, we recommend using $K>1$ periods for verification, as we do in the analysis of the racial wage gap, to which we now turn.

\section{Wage Differentials in the Federal Government}\label{Analysis}

We now illustrate how synthetic data, verification, and a PRDN could be used together to analyze pay differentials by race in the federal government.  Section \ref{wage:background} provides background on estimating pay differentials. Section \ref{wage:model} introduces our general regression modeling approach for the SF data analysis.  Section \ref{wage:overall} describes an overall analysis of average differences in pay across races, pooling all years of data.   Section \ref{wage:annual} investigates the trends in pay differentials over time.

\subsection{Prior research on pay differentials}\label{wage:background}


Social scientists have spent decades measuring the race wage gap.  Estimates based on data from the Current Population Survey, the National Longitudinal Survey of Youth, and the Panel Study of Income Dynamics, among other datasets, put the unconditional black wage gap over the past thirty years between 16\% -- 40\%.  When controlling for individual demographic and job-related variables, such as education, age, gender, occupation, and industry, estimates put the gap between 0\% -- 15\%, depending upon the precise dataset, controls, and statistical methods used \citep{AltonjiBlank1999, Cancioetal1996, CardLemieux1994,  Maxwell1994, McCall2001, NealJohnson1996, ONeill1990}.  
Estimates of the race wage gap have been declining or steady  over the past few decades \citep{Hooveretal2015, Sakano2002}.  For example, \citet{AltonjiBlank1999} estimate that the black wage gap for full time, year-round workers, controlling for education, experience, region, industry, and occupation remained steady at approximately 6.5\% from 1979 to 1995. Other studies controlling for only age, education, and location have found the gap drop from 47\% in 1940 to 18\% in 2000 \citep{Blacketal2013}.  

While most research on the race wage gap has focused on private sector labor markets, there is a steady literature measuring it in the public sector \citep{Borjas1982, Borjas1983, Kim2004, Charles2003, McCabeStream2000, Llorensetal2007, LewisNice1994}, and particularly in the federal government.   
 \citet{Lewis1998} 
found that between 1976 and 1986, the conditional race wage gap for black men declined from 21.0\% to 16.7\%, but for black women declined only from 29.7\% to 27.9\%; for Hispanic men the move was from 17.9\% to 13.0\%, and for Hispanic women it was 27.9\% to 23.3\%.
\citet{Lewis1998} also found that black men with educations and work experiences comparable to those for white men encounter a wage gap of 4\%.  He concluded that minorities made substantial progress in closing the wage gap between 1975 and 1995, especially at the very senior levels of the government.  
More recent work on the U.\ S.\ federal government found the race wage gap, controlling for demographic and agency characteristics, between 1988 and 2007 closed slightly for blacks from 7.9\% to 7.4\%, closed for Asians from 1.5\% to 0.5\%, and closed for Hispanics from 4.5\% to 2.8\% \citep[][57]{GAO2009}.  In the GAO report, the wage gaps were not broken out separately by sex.


\subsection{General modeling approach}\label{wage:model}

Following conventions in the literature on pay disparities \citep{blau:kahn}, 
we estimate the race wage gap using linear regression techniques with a standard set of demographic and human capital predictors, running the same models on both the synthetic and confidential SF datasets. The dependent variable is the natural logarithm of an employee's inflation adjusted basic pay in a given year. Basic pay is an individual's base salary and excludes any additional pay related to geographic location, award payments, or other monetary incentives paid out to employees.  
We exclude any observations with pay values of 0 or codes indicating the record is invalid, according to the OPM. 
We  use all available cases \citep{littlerubin} for regression modeling, as we have no reason to think values are systematically missing.   
In the confidential data, race is missing for 0.06\% of  person-years; 
gender is nearly always observed; education is missing for 1.82\% of person-years; age is missing for  0.01\% of person-years; occupation is missing for 0.03\% of  person-years; and agency and year are never missing.

The central independent variable is the race with which individual employees identify. Prior to 2005, employees could choose to identify with 16 categories. The largest five utilized were American Indian or Alaska Native, Asian or Pacific Islander, black, Hispanic, and white. The 
other, substantially less-utilized categories were Asian Indian, Chinese, Filipino, Guamanian, Hawaiian, Japanese, Korean, Samoan, Vietnamese, Other Asian/Pacific Islander, and Not Hispanic in Puerto Rico. We group these categories 
(save the last) with the Asian or Pacific Islander category and drop the (very small) category of Not Hispanic in Puerto Rico in accordance with government practice given the ambiguity in this category \citep[][footnote 9]{opmspringer}.
After 2005, OPM created a new combined race and ethnicity variable that enables respondents 
 to select both a race and a Hispanic ethnicity. 
Additionally, OPM collapsed the various Asian national categories into a single Asian category, and separated out Native Hawaiian and/or Other Pacific Islander into its own category. 

To make races comparable across years, we follow OPM's guidance and 
aggregate the Asian and Native Hawaiian and/or Other Pacific Islander categories to a single Asian category that is consistent with the aggregation for the pre-2005 data. Additionally, we code individuals that report a Hispanic ethnicity as Hispanic and disregard their self-reported race (if they did report one). 
In the regressions, we include indicator variables for four racial groups: American Indian/Alaska Native (AI/AN), Asian, black, and Hispanic. The omitted reference category is white.

We also include other variables plausibly correlated with race and pay.  These include the  employee's age as well as its square,
and years of educational attainment after high school.
We include fixed effects for the bureau in which an individual works to account for time-constant organizational factors that may affect wages, 
and over 800 indicators for individuals' occupations to account for  
differences in pay structures across occupations.
This is the most disaggregated occupational 
measure available.  

Previous research on the racial wage gap in the federal government 
has found substantial differences between male and female employees \citep[e.g.][]{Lewis1998}. 
We therefore 
perform analyses separately by gender. 

There is some question as to how general the occupational information included in 
regression analyses of pay disparities should be. On the one hand, 
if individuals are systematically excluded from different occupations 
on the basis of race, because of discrimination or some other factor, for instance, 
then including information on that occupation may lead 
to a biased estimate of racial pay disparities. However, at the same time,
there are important differences across occupations in terms of pay structures
and career advancement for which analysts would like to control \citep{bolton3}.  Here, we report results conditional on occupation; \citet{barrientos;etal;2017a} include results that condition only on six broad occupation classifications.

\subsection{Overall differentials}\label{wage:overall}

In the overall analysis, each observation is an employee-year.  Most individuals are observed 
for multiple years, so these observations are not independent. 
We therefore use robust standard errors that account for clustering at the employee level \citep{cameron:miller}.
We also include indicators for the year in which the observation occurs, thereby accounting for year-level shocks to wages that are experienced by all employees.

%

Mimicking the way analysts would use the integrated system, we start by estimating separate models for male and female employees using the synthetic data.  
The results in Table \ref{basic_results} reveal important relationships between race and pay in the federal government. 
In particular, according to the synthetic data results,  men who identify as AI/AN, Asian, black, and Hispanic 
are paid significantly less than comparable white male employees.  The same holds for women of all race categories except black, where
the effect is not distinguishable from zero in terms of both practical or statistical significance.  
Male (female) employees that identify as AI/AN earn approximately 0.6\% (0.9\%) less than similarly situated white male (female) employees. 
The  gaps for Asian and black male employees relative to 
white male employees appear to be significantly larger at 2.8\%  and 2.1\% percent, respectively.  
These gaps are noticeably smaller for women of these two race categories, even non-existent for black female employees.
Hispanic men and women take home about 1.4\% less than comparable white employees.




\begin{table}[t]
\begin{center}
\begin{tabular}{lrcrrrrcr}
  \hline
& \multicolumn{3}{c}{Males' Regression} & & \multicolumn{3}{c}{Females' Regression}\\ 
Variable & Synthetic & $\hat{r}$ & Confidential & &  Synthetic & $\hat{r}$ & Confidential\\
  \hline
AI/AN & -.006 (4) & .76 & -.019 (12) & &  -.009 (7) &  .97 & -.027 (19)\\
 Asian & -.028 (30) & .99 & -.040 (43) & & -.011 (13) & .42  & -.010 (11)\\
 Black & -.021 (39) & .99 & -.036 (61) & & .00013 (.3) & .003 & -.003 (8)\\
 Hispanic & -.014 (22) & .99 & -.029 (42) & & -.013 (19) & .99  & -.021 (30)\\
 Age & .033 (365) & & .043 (480) & & .023 (286) & & .032 (404) \\
 Age Sq. & -.00027 (269) & & -.00036 (352) & & -.00019 (205) & & -.00027 (295) \\
 Education & .013 (122) & & .021 (180) & & .014 (130) & & .023 (198)\\
& & & & & & & & \\
Employee-years & 13,008,298 & & 12,720,500  & & 12,263,514 & & 11,874,048\\
Employees & 1,446,499  & & 1,430,238 & & 1,390,611 & & 1,348,381\\
   \hline
\end{tabular}
\caption{Coefficients from overall regression models and posterior modes $\hat{r}$ of verification measures.   AI/AN stands for American Indian and Alaska Native, and Asian includes individuals that identify as Native Hawaiian or Pacific Islander.  Absolute values of $t$-statistics are in parentheses.
Disparities in sample sizes arise from deletions of cases with missing values in the confidential data analyses.\label{basic_results}}
\end{center}
\end{table}

The analyst next would submit requests for verification of these results.
For each race coefficient in Table \ref{basic_results},
we make a separate verification query using the method in \eqref{bayes_model} with $\epsilon = 1$.
We group employees into  $m=50$ partitions, so that each employee is a member of only one partition.  
Thus, in the language of $\epsilon$-DP, 
the neighboring databases differ in one employee, as opposed to one 
employee-year observation.  The former is more sensible for verifications of the overall regression.   
A data snooper with knowledge of all but one employee-year observation  could figure out many, if not all, of
the values for the missing observation by logical deduction, e.g., easily inferring the age of the missing year and 
bounding the salary between the previous and successive years. 
We set the threshold $\gamma_0 = -.01$, and target queries at whether $\beta_j < -.01$ or not.

As evident in Table \ref{basic_results}, the posterior modes of the verification measure clearly 
indicate that the wage gaps for male employees who are black, Asian, and Hispanic are all at least 1\%. 
The evidence of at least a 1\% wage gap for AI/AN men is less obvious but still suggestive, with a posterior mode around .75. 
Thus, the verification measures validate the findings from the synthetic data regressions of 
substantial racial wage gaps for black male, Asian male, and Hispanic male employees, and they suggest the synthetic
data results for AI/AN male employees are close to accurate as -.006 is not far from -0.01.  For women, the posterior models of 
the verification measure clearly indicate at least 1\%  wage gaps for Hispanic and AI/AN employees.
They also provide strong evidence against a wage gap of at least 1\% for female black employees, 
with a posterior mode near zero. For female Asian employees the verification measure suggests 
the wage gap could be almost equally likely above or below 1\%, as the posterior mode equals .42.  This 
suggests that the true coefficient is likely near -0.01.
Thus, the verification measures validate the findings from the synthetic data that the wage gap for Hispanic female
employees is at least 1\%, but that there is not a substantial wage gap for black female employees.  They also suggest that the
estimate for AI/AN (-.009) could be an underestimate, since the verification measures suggest that the true coefficient
 is indeed less than -.01.  Finally, they suggest that the synthetic data coefficient for female Asian employees 
is likely accurate, since it is close to -.01.

We expect that some users might be satisfied with this level of verification, and thus can publish the synthetic data results plus the verification answers.
However, others may want to perform the analysis on the confidential data via the remote access component of the system.  As shown in 
Table \ref{basic_results}, in the confidential data regression the estimated coefficients for all 
four race indicators are negative and statistically significant for both genders, suggesting that non-white employees earn
less than white employees. The estimated gaps for men are at least 1.9\% across races, with particularly large gaps 
for black men (3.6\%) and Asian men (4.0\%). For women, AI/AN, Asian, and Hispanic employees earn 2.7\%, 1.0\%, and 2.1\% less than comparable white female workers. Strikingly, the coefficient estimate for black women is essentially zero, suggesting parity with similarly situated white women.  The wage gaps for black women and Asian women are substantially smaller than for men of those race categories.


The effect sizes from the confidential data are fairly similar 
to those from the synthetic data.  This is in accord with the conclusions from the verification measure.   The most practically relevant difference in the synthetic and confidential data results
 exists for employees that identify as AI/AN: the synthetic data show gaps of less than 1\% whereas the confidential data
show gaps of at least 1.9\%. This group of employees is the smallest racial group in the federal 
government, making it challenging to create accurate synthetic data for them.


\subsection{Year-by-year results}\label{wage:annual}

\begin{figure}[t]
\centering
\subfigure{\includegraphics[scale = 0.44]{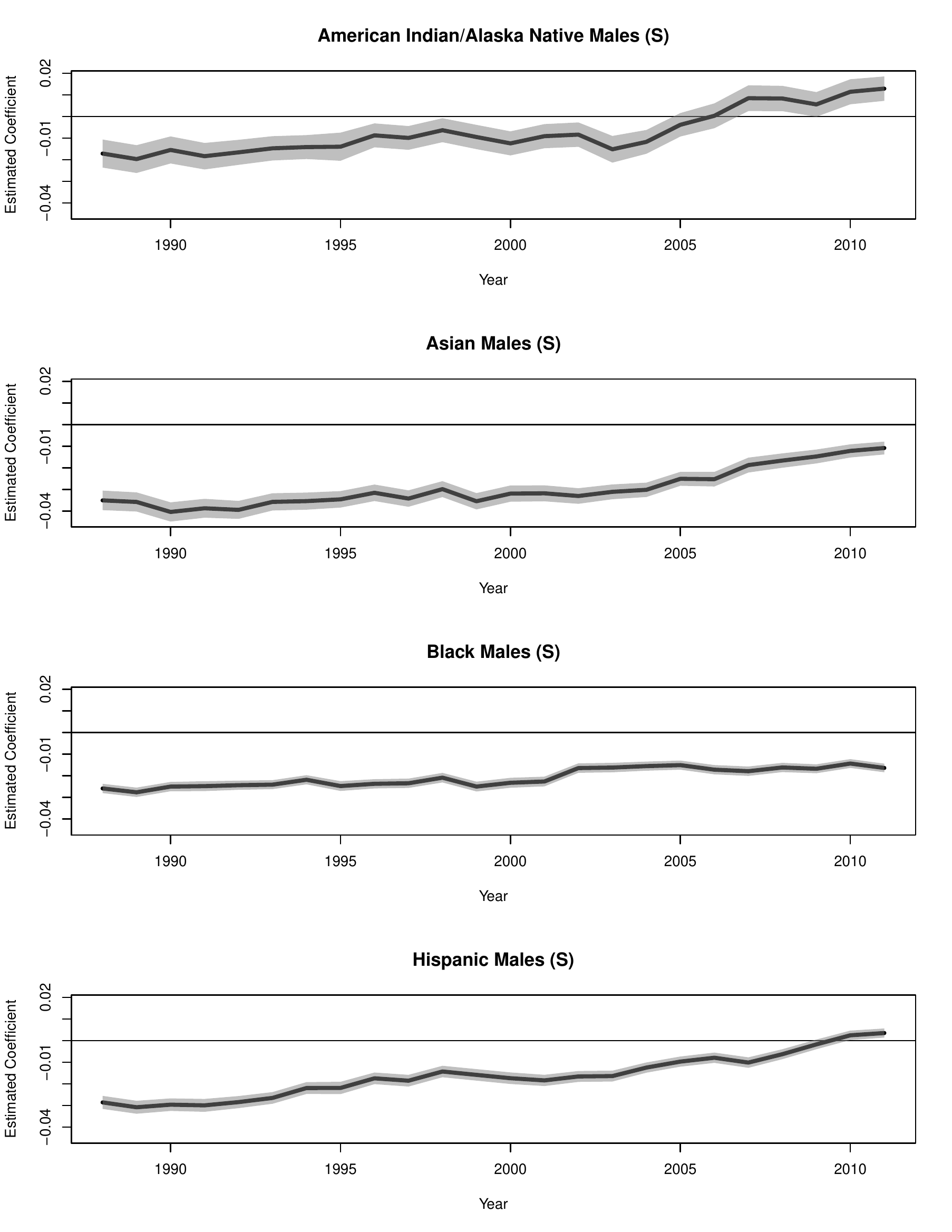}}
\subfigure{\includegraphics[scale = 0.44]{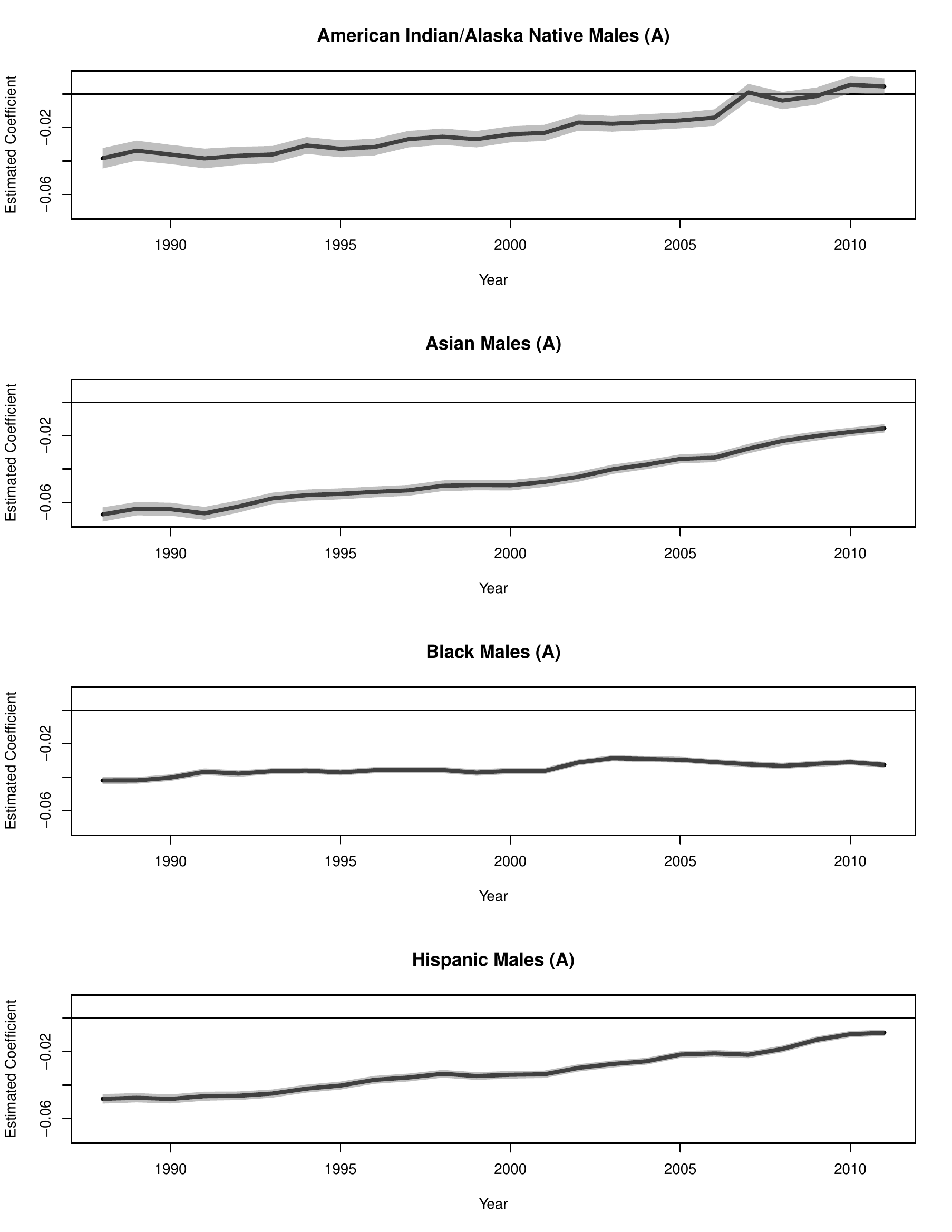}}
\caption{Estimated racial wage gaps (coefficients of race indicators) for yearly males' regressions in synthetic data (left) and confidential, authentic data (right).}
\label{synthetic_trend_m}
\end{figure}

We next turn to year-by-year estimates of pay gaps in order to examine potential trends over time. 
We estimate the same models used in Table \ref{basic_results},
except run on each year of data separately. As before, we start with the synthetic data.
The synthetic data results in Figure \ref{synthetic_trend_m} suggest that 
the wage gap for men has shrunk steadily over the period of the study in all race groups but black males.
For black males, the estimated gap appears to be relatively stable throughout the time period.   By 2011 in the synthetic data, the wage gap appears to have disappeared for AI/AN and Hispanic men, and reduced to around -1\% for Asian men.

\begin{figure}[t]
\centering
\subfigure{\includegraphics[scale = 0.44]{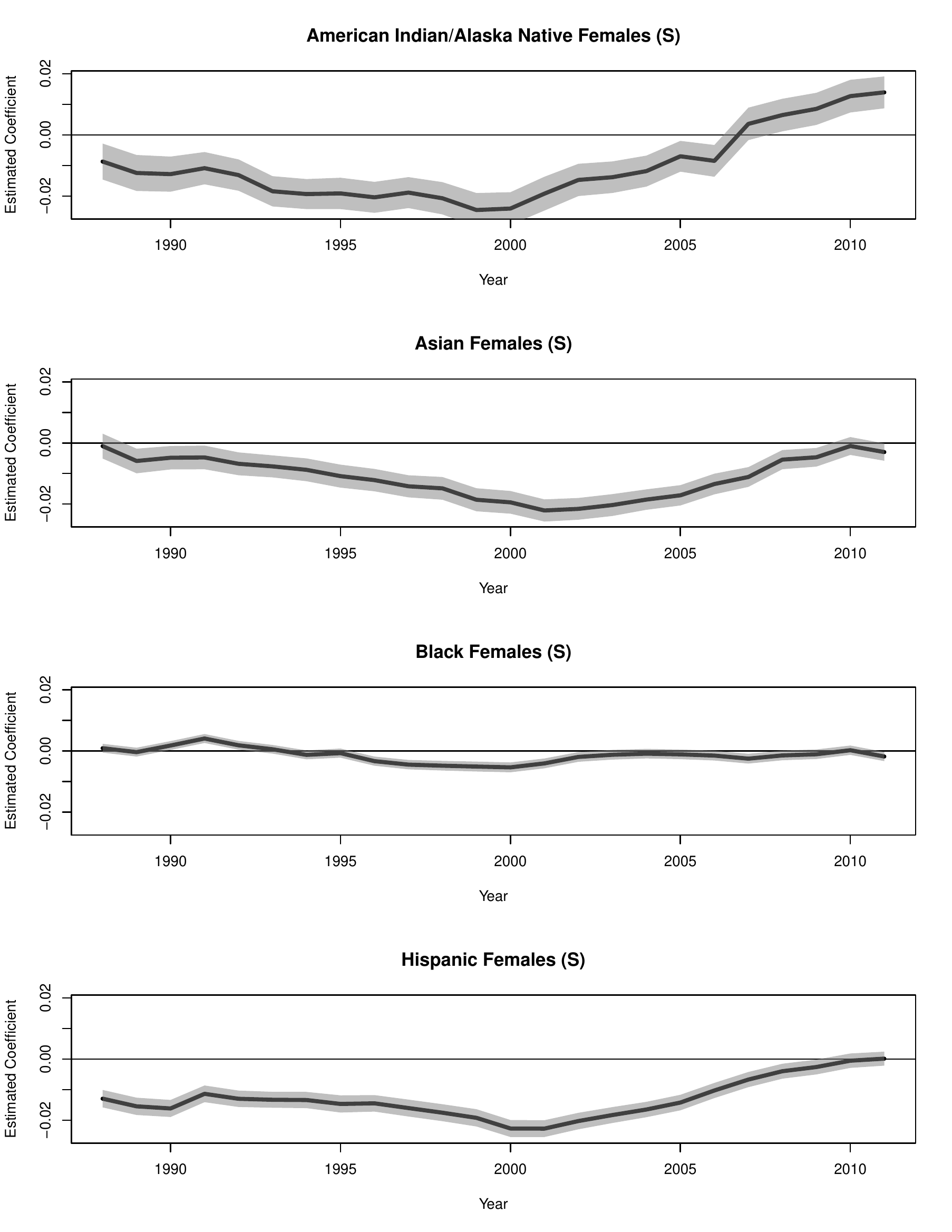}}
\subfigure{\includegraphics[scale = 0.44]{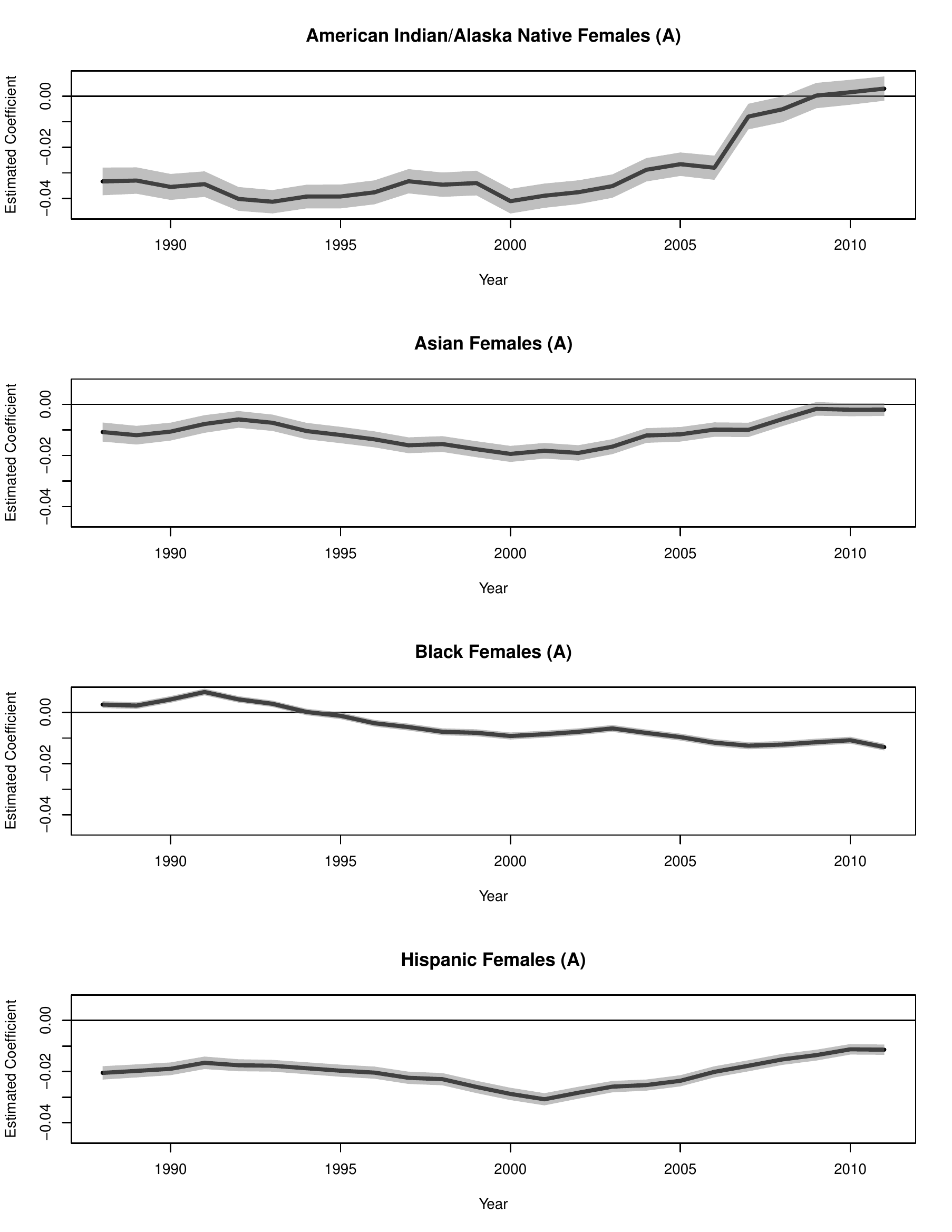}}
\caption{Estimated racial wage gaps (coefficients of race indicators) for yearly females' regressions in synthetic data (left) and confidential, authentic data (right).}
\label{synthetic_trend_f}
\end{figure}

For female employees, the story from the synthetic data  is more complicated.
Figure \ref{synthetic_trend_f} suggests that 
AI/AN, Asian, and Hispanic women all had declining wages relative to white women until the early 2000s, when the trend largely reversed, with all three groups making progress toward parity. Indeed, the synthetic results indicate that AI/AN women actually earned more than comparable white women after 2006. For black women, the synthetic data estimates of the wage gap change only slightly, from 0.1\% in 1988 to -0.2\% in 2011, suggesting negligible wage gaps at any time point in time during the study.


To verify these trends, we estimate the longitudinal measure described in Section \ref{sec:ver2}.
Looking at Figure \ref{synthetic_trend_m} for male employees, analysts might consider two sets of time periods $\{\mathcal{T}_k\}$.
The first is an overall trend, setting $K=1$ and $\mathcal{T}_1 = \mathcal{T}$ for all races.  The second uses $K=2$ periods, with a bifurcation at
a year where the pattern deviates most noticeably. These are the years 2003 for AI/AN men, 2002 for Asian men, 1998 for black men, and 1997 for Hispanic men.
We do the same for female employees, using Figure \ref{synthetic_trend_f} to identify bifurcations at 1998 for AI/AN women, and at 2000 for all other female employees. 
We set each $C_k$ to indicate whether the slope is positive $(C_k = [0, \infty])$ or negative $(C_k = [-\infty, 0])$.  Of course, one could examine 
other time periods and intervals.  For each verification, we use $\epsilon = 1$ and $M=50$ partitions, ensuring that each employee appears only once in each partition.  

\begin{table}[t]
\begin{center}
\begin{tabular}{lccccc}
  \hline
& \multicolumn{2}{c}{Males} & & \multicolumn{2}{c}{Females}\\ 
Coefficient & Interval & $\hat{r}$ & & Interval & $\hat{r}$ \\
  \hline
AI/AN &      1988 - 2011   &     .94 & & 1988 - 2011  &       .94 \\
       &             1988 - 2003  &      .82 &  & 1988 - 1998  &       .60 \\

        &            2003 - 2011    &    .91 & & 1998 - 2011    &    .98 \\

 Asian  &     1988 - 2011   &     .99  & & 1988 - 2011    &    .33 \\

&                    1988 - 2002  &      .89  &  & 1988 - 2000  &      .72 \\

&                    2002 - 2011     &   .96  & & 2000 - 2011     &   .98 \\

  Black  &     1988 - 2011   &     .89  &  & 1988 - 2011     &   .99 \\

&                    1988 - 1998   &     .74 &  &  1988 - 2000   &      .98 \\
 
&                    1998 - 2011     &   .71 &  & 2000 - 2011   &     .14 \\

      Hispanic  &  1988 - 2011   &     .99  & & 1988 - 2011    &    .55 \\

&                    1988 - 1997     &   .85 &  & 1988 - 2000    &    .74 \\

&                    1997 - 2011      &  .99  & &  2000 - 2011   &     .97 \\
\hline
\end{tabular}
\caption{Posterior modes $\hat{r}$ of verification measures for year-by-year trends.   AI/AN stands for American Indian and Alaska Native, and Asian includes individuals that identify as Native Hawaiian or Pacific Islander.\label{tab:ver:yearly}}
\end{center}
\end{table}

Table \ref{tab:ver:yearly} displays the posterior modes of the verification measures for the two sets of periods.  
For men, the posterior modes are all at least 0.7 for all time periods and all races, with most above 
0.9.  This indicates that the trends in the synthetic data coefficients accord well with the trends in the confidential data regressions   
for these two sets of periods.  For women, however, the verification results give reason to doubt some of the trends in the synthetic data regressions. For AI/AN women, we see strong agreement in the overall trend over all years and the trend from 1998 onward, 
but some uncertainty 
about the trend between 1988 and 1998.  Verification values around .50 are consistent with a nearly flat trend in the confidential data coefficients, 
which is also the trend in the synthetic data. For Asian women, we see strong agreement in the synthetic and confidential 
regression trends over 2000 to 2011, modest agreement from 1988 to 2000, and poor agreement over the whole period.  The 
synthetic data trend suggests the wage gap for Asian women in 2011 is nearly the same value as in 1988; however, the verification 
measures suggest that is not the case. 
For black women, the verification results confirm that the wage gap increased over the 24 years as a whole, and
in particular between 1988 and 2000. However, the trend in the synthetic data coefficients---which 
suggests black women actually caught up to white women---is not accurate.  With a posterior mean of .14, we clearly should not trust the 
trend for black women in the synthetic data after 2000.  For Hispanic women, we see strong agreement in the synthetic and 
confidential regression trends after 2000, and modest agreement from 1988 to 2000.  Between 1988 and 2011, however, the verification measure
 is close to 0.5, suggesting that the trend line from the confidential data is nearly flat for Hispanic women.

Turning to the results on the confidential data, 
Figures \ref{synthetic_trend_m} and \ref{synthetic_trend_f} show that the race wage gap has been shrinking for all groups except black female employees.   
In the confidential data, we estimate a significant decline in the position of black women relative to white women in the federal service during the time period of our study. In 1988, we estimate that black women earned 0.3\% \emph{more} than similar white counterparts. By 2011, black women were earning approximately 1.4\% \emph{less} than white women with similar demographics and occupations. 

The trends observed in the synthetic dataset are largely mirrored in the confidential data, with the exception of black female employees. This was apparent in the verification measures, as well, 
which highlighted the mismatch in the trends for black women after 2000.  In both analyses, however, it is clear that black women have not experienced the gains that women identifying with other races have relative to white women. 
In general, estimated coefficients from the synthetic data analyses tend to be smaller in magnitude than those from the confidential data analyses. There are some sign discrepancies for the estimated coefficients as well. 
For instance, in the synthetic results, Hispanic males are estimated to have higher levels of pay relative to comparable white men in the final years of the analysis. 
However, the coefficient estimates from the confidential data for Hispanic males never exceed zero.

\section{Concluding Remarks}\label{Conclusion}

The integrated system described here enables the work flow illustrated by the OPM analysis: start with synthetic data, verify results, and access the confidential data via a PRDN when necessary. Although one could implement 
systems that exclude some of the components, integrating all three components has advantages.  We now describe some of these benefits, framing the discussion around examples of alternative systems that exclude one or more of the components.

One possibility is not to bother releasing record-level synthetic data at all. Instead, the data steward 
only allows users to query a system for disclosure-protected outputs of analyses.
While in some contexts providing only outputs may be sufficient, we believe that providing 
access to record-level synthetic data has enormous benefits.  
Record-level data provide readily accessible testbeds for methodological researchers to evaluate their latest techniques.  
They help students and trainees, who may not be able to gain approval to use a PRDN or other secure data enclave, learn the skills of data analysis.  
Even for experienced researchers, large-scale data can be difficult to ``get your head around'' because of complexities and structural subtleties that are difficult to learn without seeing the data. 
Researchers often do not know in advance which are the right questions to ask of the data or the best modeling choices for addressing those questions.  
As noted by \citet{karr:reiter:14}, exploratory analyses dealing with the data themselves are a fruitful path to the right questions.  

Another possibility is to skip the verification step and allow for direct validation of results.  That is, allow the user to send code implementing the analysis of interest to the data holder, who runs the code on the confidential data and returns disclosure-protected output to the user.  In fact, the Census Bureau has this system in place, offering validation of results for some of its synthetic data products \citep{vilhuberabowdreiter}.  The disclosure protection involves {\em ad hoc} methods like rounding estimates to a small number of significant digits and ensuring estimates are based on some minimum number of cases.  Validation as implemented currently and differentially private verification have similar goals, but there are differences.   Verification servers can provide immediate, automated feedback on the quality of analysis results, whereas validation typically requires some amount of manual labor on the part of the data holder, who must investigate the code and the outputs for disclosures, decide what treatments to apply, and repeat as necessary.  Done once this checking may not be too onerous, but done repeatedly it can cost the user and the data holder significant time, depending on the resources the data holder makes available for validation.  
Additionally, with differentially private verification the data holder can track the total privacy loss of any sequence of verification requests, which provides a provable bound on the disclosure risk from allowing comparisons of the real and synthetic data results; this is not the case with repeated applications of 
 {\em ad hoc} validation methods.
On the other hand, for users who want results from the confidential data for a specific analysis regardless of what a verification would reveal, skipping the verification step could result in fewer queries of the confidential data, which has advantages for privacy protection.  

A third possibility is to allow verification of synthetic data results but not provide users with results from the confidential data via a PRDN or  validation service.  This may be fine for users whose analyses are faithfully preserved in the synthetic data, but this is clearly inadequate for other users.  As seen with the OPM application and in other synthetic datasets \citep{abowd06, lbdisr}, inevitably synthetic data are unreliable for some analyses.  

Validation is implicitly a central part of the integrated system outlined here. Outputs from approved researchers' analyses of the confidential data via the PRDN still should be subject to disclosure control treatment.  Ideally, these treatments satisfy $\epsilon$-DP, so that the system can keep track of the total privacy budget spent on verification and validation of results. Indeed, an interesting research topic is to optimize the usage of differentially private verification and validation for a fixed privacy budget. Differentially private output perturbation is available for many types of analyses, especially those based on counts \citep{dwork:roth}. We are not aware of differentially private regression output perturbation techniques that scale with low error to models of the dimension and complexity used here, i.e., a data matrix with 28 million rows and over 800 columns. Most existing methods for differentially private regression generate only point estimates of coefficients without standard errors; some do so under restrictions on the sample spaces of the dependent or independent variables not relevant for the OPM data.   Given the current state of the art, we suggest that data holders  use $\epsilon$-DP methods when available and use {\em ad hoc} methods currently employed by statistical agencies when not.
This points to two other important research topics, (i) developing and testing the practical performance of formally private regression algorithms for large dimensional models and data, and (ii) assessing disclosure risks inherent in coupling {\em ad hoc} disclosure treatment of outputs with differentially private verification measures. 

This last research question can be extended to the entire integrated system: how do we characterize the disclosure risks inherent in releasing synthetic data like the kind we generated for the OPM data, coupled with $\epsilon$-DP verification or validation measures, and {\em ad hoc} validation measures when needed?  We do not have a general answer to this question, as it presents difficult conceptual and computational challenges; this is a topic for future research.  One goal to work toward is to require all validations and all synthetic data generators to satisfy $\epsilon$-DP, and rely on composition properties to provide bounds on the risk.  To our knowledge, this currently is not possible (at acceptably low levels of error) with state-of-the-art techniques for generating $\epsilon$-DP synthetic data \citep[e.g., ][]{baraketal, abowdvilh,  blumetal, onthemap, annesophie, hardt,  miretal, karwa:sesa:annals}  for data with the dimensionality and complexity of the OPM data. However, work is ongoing.  
Because the integrated system is inherently modular, one can substitute other techniques for synthetic data generation, verification, and validation appropriate for the tasks at hand.

Coupling synthetic data with verification and validation services has benefits for data stewards as well. In particular, because analysts have opportunities to verify and validate results, it is not overly problematic if the synthetic data provide low quality answers for some analyses.  This leeway can allow data stewards to use relatively straightforward modeling strategies with automated fitting routines, like those we used to generate the OPM synthetic data, in place of comprehensive or computationally expensive modeling strategies.  Related, it also can allow  data stewards to use synthesis routines that result in greater privacy protections at the cost of lower analytical validity.

The $\epsilon$-DP verification measures are based on binary variables computed on sub-samples of the confidential data.
This is a generic method that can be adapted to handle comparisons for many types of models, making it a flexible strategy for verification.  However, it is not always effective for analyses based on small samples, as documented in \citet{barrientos;etal;2017a}. In fact, for some analyses and datasets, 
the partitioning process can result in inestimable regressions.  For example, the random sub-sampling may result in partitions that have perfect co-linearities or dummy variables with 
all values equal to zero. Many software packages automatically drop such variables and  report coefficients for the remaining variables, making it still possible to compute the measures although
complicating interpretations of the results.  When errors make it impossible to obtain results, we suggest adapting 
the binary measure by adding a third category of counts corresponding to the number of errors. Here, the outputs of the measure 
include the number of ones, zeros, and errors.  We can protect these counts using the Laplace Mechanism, and report posterior
modes of the number of errors and the fraction of ones among cases without errors.  \citet{barrientos;etal;2017a} present and evaluate this variant of the algorithm.  
In the wage gap analyses reported here, fitting errors did not occur due to the large sample sizes.

The choice of the number of partitions is up to the data analyst; we used $M=50$ in the wage gap  verification.  Analysts should strive to make $M$ as large 
as possible to minimize the impact of the Laplace noise on the verification counts.  On the other hand, users should  
allow $r$ to be as close to one (or zero) as possible, as these values are easiest to interpret.  Making $M$ too 
large flattens the distribution of the MLEs in the partitions, thereby moving $r$ toward 0.5 and more uncertain verification decisions.  
We found that $M=50$ gave a satisfactory trade off for the analyses presented in Section \ref{Analysis}.   We recommend that analysts experiment with the synthetic data to 
find a suitable $M$ for their analysis of interest.    Another possibility is to spend some of the privacy budget on selecting an optimal $M$ 
from a discrete set of choices, according to some loss function that depends on $M$ and $r$.  Developing such measures is an area for future research.

Using the verification server, or any differentially private data release strategy, requires the data steward to set privacy budgets.  \citet{abowdschmutte} describe the data steward's task of setting $\epsilon$ as balancing trade offs in the amount of privacy protection and degree of data usefulness desired by society for the data product at hand.  They use survey data to assess people's preferences for privacy and a loss function on the accuracy of estimates (of an income distribution) to express data usefulness.  To date, however, there have been few examples of differentially private data releases in genuine production settings, and hence few practical guidelines for setting $\epsilon$. The Census Bureau   releases differentially private synthetic data showing maps of the street blocks where people live and work based on $\epsilon \approx 9$ \citep{onthemap}. The data are released once every year, and a new privacy budget (of about 9) is used for each release. Google Chrome uses a differentially private algorithm called RAPPOR \citep{RAPPOR} to collect sensitive browser characteristics from users continuously (every day). They use a per-day $\epsilon= log(3)$. Recent work \citep{appleDP} has analyzed the algorithms used by Apple MacOS Sierra (Version 10.12) to collect sensitive user data that have been claimed to satisfy differential privacy. Through a combination of experiments, as well as static and dynamic code analysis, this work estimates the privacy loss per each data item sent to Apple's servers to be $\epsilon = 1$ or $\epsilon = 2$, and the overall privacy loss to be around 16 per day. 


Regardless of the choice of $\epsilon$, if the data steward follows $\epsilon$-DP strictly, at some point the total privacy budget allowed will be exhausted, at which point no new analysis results may be released.  
With a finite privacy budget, data stewards have to decide who gets access to the system and in what order.  These raise complicated issues of fairness and 
evaluation of the importance of analyses, which have yet to be addressed in production settings with interactive query systems.  The privacy budget used by a sequence of verification requests is an instance of a more general problem of optimizing the privacy loss of a sequence of queries over the database  
\citep[e.g.,][]{hardt:rothblum, ullman}. However, existing work only considers optimizing the privacy budget for a sequence of a special class of aggregation queries called linear queries (e.g., counts and histograms). Optimizing the privacy budget for an adaptively chosen sequence of complex tasks (e.g., regression or verification) is an open research problem. 
 

For integrated systems that utilize $\epsilon$-DP, it seems likely that any reasonable overall privacy budget will be exhausted quicker than desired. We see two general paths to addressing  this dilemma.  One is  technological: it should be possible to develop verification measures that use less of the privacy budget.  For example, our verification measures operate independently on coefficients, but there may be ways to leverage correlation among the coefficients to do multivariate verification. For specific analyses and verification tasks, it may be fruitful to work directly with the coefficients rather than through methods based on sub-sampling.   

Another approach is policy-oriented: data stewards can give up some of the global protection from differential privacy to enhance access. 
For example, rather than enforce an overall privacy budget over all queries of the data, the data steward can provide individual, approved users with a finite privacy budget. 
In this case, the data steward trusts users not to collude with each other to circumvent the protection offered by the formal privacy.  This type of policy is common in other privacy preserving data analyses, such as secure multi-party computation \citep{techno06}
Alternatively, and the approach implicitly taken in our illustrative application of synthesis plus verification, 
the data steward can allocate a privacy budget per analysis.  In this case, the data steward trusts individual users not to attempt to learn sensitive information from repeated queries. 
With these compromises, the data steward can control the information leakage from providing verifications for any individual user or analysis, while still offering the convenience and usefulness of automated verification. 
 When coupled with penalties for malfeasance and other policies, the data steward might consider either compromise reasonable, as legitimate researchers generally are interested in scientific inference rather than attacking privacy protections.  

Finally, we
are developing a verification server and associated $R$ package that implements the verification 
measures from Section \ref{verification}.  This system allows analysts to run verification measures on synthetic data, so that they can assess the likely usefulness of the measures for their analysis of interest, as well as select effective values of $M$. The package also offers methods for generating $\epsilon$-DP  plots of residuals versus predicted values, thereby helping users assess the reasonableness of the assumptions of a posited model when applied on the confidential data \citep{icdm16paper}. 
The codebase for the verification measures, as well as code used to generate the synthetic OPM data, are available in \citet{barrientos;etal;2017c}.  
We intend to make the package available on CRAN, so that data stewards and other researchers can experiment with and further develop this framework for providing access to confidential social science data.

\section*{Supplementary Material}

{\bf Supplement A to Providing Access to Confidential Research Data Through Synthesis and Verification: An Application to Data on Employees of the U.S.\  Federal Government} (DOI: ; .pdf). This document provides supporting material for aspects of the OPM synthesis plus verification application. In Section 1, we provide a formal description of the three sub-models used to model the employee's career. In Section 2, we discuss the modeling strategies used to synthesize variables in the OPM data.  In Section 3, we provide the full list of the synthesized variables along with a brief description of each of them.  In Section 4, we present the analyses of wage gaps conditional on six broad categories of occupation rather than the 803 used in the main text. In Section 5, we describe a method for empirical disclosure risk assessment for OPM synthetic data. In Section 6, we formally describe the verification measures for longitudinal trends in regression coefficients.  In Section 7, we examine the performance of the $\epsilon$-differentially private verification measures used in the text, and we present a verification measure that is suitable for analyses where some regression coefficients are nonestimable

{\bf Supplement B to Providing Access to Confidential Research Data Through Synthesis and Verification: An Application to Data on Employees of the U.S.\  Federal Government} (DOI: ; .pdf).  This document provides graphical analyses comparing the OPM synthetic and confidential data used in the main text.

{\bf Supplement C to Providing Access to Confidential Research Data Through Synthesis and Verification: An Application to Data on Employees of the U.S.\  Federal Government} (DOI: ; .zip). This file contains the code used to generate the synthetic OPM data and compute the verification measures proposed in the main text.



\end{document}